\documentclass[letterpaper, 10 pt, conference]{ieeeconf}  

\IEEEoverridecommandlockouts                              

\usepackage{cite}
\usepackage{amsmath,amssymb,amsfonts}
\usepackage{graphicx}
\usepackage{textcomp}
\usepackage[dvipsnames]{xcolor}
\def\BibTeX{{\rm B\kern-.05em{\sc i\kern-.025em b}\kern-.08em
    T\kern-.1667em\lower.7ex\hbox{E}\kern-.125emX}}

\usepackage[final]{changes}

\usepackage{comment}

\usepackage{mleftright}

\usepackage{import}

\usepackage{algorithm}
\usepackage{algpseudocode}

\usepackage{pgfplots}
  \usepgfplotslibrary{groupplots}
  \usepgfplotslibrary{fillbetween} 
\pgfplotsset{compat=newest}

\usepackage{xurl}
\usepackage[hidelinks]{hyperref}

\usepackage{subcaption}

\usepackage{tikz}
\usetikzlibrary{positioning, arrows, calc, shadings}
\usetikzlibrary{fit}

\tikzstyle{blockdiag}	= [node distance=5mm, >=stealth', semithick]
\tikzstyle{block}			= [draw, rectangle, minimum width=1cm, minimum 
height=.8cm]
\tikzstyle{sum} = [draw,circle,inner sep=0pt, minimum size=6pt]
\tikzstyle{connector} = [draw,circle,inner sep=0pt, minimum size=2pt, 
fill=black]

\usepackage{array}
\newcommand{\norm}[1]{\left\|#1\right\|}
\newcommand{\abs}[1]{\left|#1\right|}
\newcommand{\field}[1]{\mathbb{#1}}
\newcommand{\R}{\field{R}}

\newcommand{\bmtx}{\begin{bmatrix}}
\newcommand{\emtx}{\end{bmatrix}}
\newcommand{\bsmtx}{\left[ \begin{smallmatrix}} 
\newcommand{\esmtx}{\end{smallmatrix} \right]} 
\newcommand{\bmatarray}[1]{\left[\begin{array}{#1}}
\newcommand{\ematarray}{\end{array}\right]} 

\newcommand{\mcl}[1]{\mathcal{#1}}

\newcommand{\vspaceeq}{\vspace{-3pt}}

\definecolor{blue1}{RGB}{222,235,247}
\definecolor{blue2}{RGB}{158,202,225}
\definecolor{blue3}{RGB}{49,130,189}

\newcommand{\Dx}{\mathcal{D}}
\newcommand{\U}{\mathcal{U}}

\DeclareMathOperator*{\argmin}{arg\,min}

\newtheorem{mytheo}{Theorem}

\newtheorem{defn}{Definition}
\newtheorem{remark}{Remark}
\newtheorem{cor}{Corollary}

\newcommand{\edtn}[1]{\null} 

\title{\LARGE \bf
Robust Time-Varying Control Barrier Functions \\
with Sector-Bounded Nonlinearities*
}

\author{Felix Biert\"umpfel, Jungbae Chun and Peter Seiler
\thanks{*This work was supported by the EU under Grant No. 101153910.}
\thanks{F. Biert\"umpfel is with the Chair of Flight Mechanics and Control at TU Dresden, Dresden, and the Department of Electrical and Computer Engineering at the University of Michigan, Ann Arbor  {\tt\small felix.biertuempfel@tu-dresden.de}.}%
\thanks{J. Chun and P. Seiler are with the Department of Electrical and Computer Engineering at the University of Michigan, Ann Arbor, {\tt\small jungbaec@umich.edu}, {\tt\small pseiler@umich.edu}}%
}

\begin{document}

\maketitle
\thispagestyle{empty}
\pagestyle{empty}

\begin{abstract}

This paper presents a novel approach for ensuring safe operation of systems subject to input nonlinearities and time-varying safety constraints. 
We extend the time-varying barrier function framework to address time-varying safety constraints and explicitly account for control-dependent nonlinearities at the plant input. 
Guaranteed bounds on the input-output behavior of these nonlinearities are provided through pointwise-in-time quadratic constraints. The result is a class of robust time-varying control barrier functions that define a safety filter. This filter ensures robust safety for all admissible nonlinearities while minimally modifying the command generated by a baseline controller. We derive a second-order cone program (SOCP) to compute this safety filter online and provide \added{novel} feasibility conditions for ball-constrained inputs. The proposed approach is demonstrated on a spacecraft docking maneuver.
\end{abstract}

\section{Introducton}\label{sec:intro}
\vspace{-5pt}

Engineering problems must meet stringent safety requirements in uncertain and time-varying environments. For example, a driver assistance system must avoid unexpected obstacles and spacecraft need to avoid late-detected space debris during maneuvering. However, these applications usually have well-established control design and verification pipelines resulting in reluctance to perform radical design changes. Thus, novel safety features should minimally alter the original design while showing robustness to comply with certification standards.

Control barrier functions (CBFs) \cite{Ames2017, ames2019control} are one popular approach to address these issues. \replaced{They are used to formulate online safety filters that guarantee safety by slightly adjusting baseline control commands through quadratic programs.}{These are online safety filters that guarantee safety by slightly adjust baseline control commands through quadratic programs.} CBFs are naturally defined for systems with relative degree one. Higher degrees are handled by exponential control barrier functions (ECBFs) \cite{nguyen2016exponential} and higher-order control barrier functions \cite{xiao2019control, tan2021high, wang2023high}. Time-varying (exponential) barrier functions (TVCBFs) \cite{xu2018constrained, lindemann2018control} address time-varying safe sets.

Existing CBF methods rely on either system models \cite{ames2019control} or data-driven approaches \cite{taylor20a, Choi2020, Bajelani2025}.
To handle uncertainties and disturbances, robust control barrier functions have been proposed to ensure robust safety  with respect to time-invariant safe sets \cite{blanchini2015set}. The approach in \cite{buch2021robust} leverages quadratic constraints to handle unmodeled input nonlinearities. A generalization to integral quadratic constraints in \cite{seiler2021control} allows for uncertainties with memory. 
Other formulations address vector field perturbations with $L_\infty$ disturbances \cite{Nguyen2022}, or apply mixed-monotone decompositions and robust optimization \cite{pati2023robust}. Additional robustness schemes include \cite{choi2021robust, safari2024time, hamdipoor2023safe}, with recent links to predictive control \cite{huang2025predictive}. 
Robust CBF approaches have been extended to time-varying safe sets to address norm-bounded disturbances \cite{Tezuka2022} or estimated parametric uncertainties \cite{Kim2025}
However, robust methods that simultaneously account for time-varying safety constraints and input nonlinearities have not yet been explored.

\replaced{The first key}{The} contribution of our paper is to address this gap using a novel robust time-varying control barrier function (RTVCBF). The proposed framework, introduced  in Section~\ref{sec:rtvcbf}, extends time-varying barrier function methods~\cite{xu2018constrained, lindemann2018control} (Section~\ref{ss:tvcbf}) to systems with unknown nonlinearities at the plant input while ensuring safety with respect to time-varying safe sets. In particular, we employ pointwise-in-time quadratic constraints to bound the input-output behavior of the nonlinearities by leveraging \added{and generalizing} derivations in \cite{buch2021robust}. This is summarized in Section~\ref{ss:rcbf}. We use this to derive robust safety conditions for time-varying safe sets that hold for all admissible nonlinearities.  We construct these conditions explicitly for systems with relative degree two.  This is an important class in practice and the results can be easily generalized, with only notational changes, to systems with relative degree different from two. The robustly safe control actions are computed from a second-order cone program (SOCP). 
\added{Our results also hold for parametric uncertainties at the plant input.} 

\added{The second key contribution, given in Section~\ref{ss:feas}, is the derivation of novel feasibility conditions under uncertainty and unit-norm control inputs.} \added{Recent work on CBF feasibility focuses mostly on offline methods for nominal systems. This includes set-wise compatibility certification via offline LP \cite{mousavi2025vertices}, exact feasibility domains for LTI systems with affine CBFs \cite{mousavi2026structure}, forward-in-time feasibility via a single auxiliary CBF constraint \cite{xiao2022sufficient}, and offline SoS-based verification handling uncertainty through set-valued dynamics \cite{isaly2024feasibility}. In contrast, we provide a novel feasibility test that can be efficiently performed online prior to solving the SOCP for the safe input.
} Finally, the proposed framework is illustrated in Section~\ref{sec:exmp} through a CubeSat docking example.


\vspace{-3pt}
\section{Preliminaries}\label{sec:pre}
\vspace{-5pt}

\subsection{Problem Formulation}

We consider the feedback interconnection of a controller $k$ and an uncertain system $G_\phi$ shown in Fig.~\ref{fig:CLIC}. The uncertain plant consists of the interconnection of a known nonlinear, input-affine system $G$ and an unknown nonlinearity $\phi$:
\begin{equation}
\label{eq:plantG}
\begin{split}
    \dot{x}(t) & = f(x(t)) + g(x(t)) \, \left( u(t)+w(t) \right), \\
    w(t) & = \phi( u(t), t ),
\end{split}
\end{equation}
where $x(t)\in \Dx \subseteq \R^n$ and $u(t)\in \U \subseteq \R^m$ denote the state and input at time $t$, respectively.  The functions $f: \Dx \rightarrow \R^n$ and $g:\Dx \rightarrow \R^{n\times m}$ are locally Lipschitz continuous. 
The admissible sets of states and inputs  are denoted by $\Dx$ and $\U$.

\begin{figure}[h!]  
 \centering
 \begin{tikzpicture}[blockdiag]
	
	\node[block](Plant){$G$}; 
	\node[sum, left = of Plant, xshift = +0.2cm](Sum2){};
	\node[block,above = of Sum2, xshift = -0.8cm, yshift = -0.30cm](Phi) {$\phi$}; 
    
    \node[connector, left = of Sum2, xshift = -1.0cm](Con1){};
	
    \node[block, left = of Con1, xshift = -0.2cm](SF){$F$};
	\node[block, left = of SF, xshift = +0.05cm](C){$k_0$};
	


\draw[->](Sum2.east) -- (Plant.west);
\draw[-](SF.east) -- (Con1.west) node[pos=0.5,above] {$u$};
\draw[->](C.east) -- (SF.west)node[pos=0.5,above] {$u_0$};
\draw[->](Con1.north) |- (Phi.west);
\draw[->](Phi.east) -| (Sum2.north) node[pos=0.3,above] {$w$};
\draw[->](Con1.east) -- (Sum2.west);



 \node[draw, dashed, inner sep=3pt, fit=(C)(SF), label=above:{Controller, $k$}] (frame) {};
  
 \node[draw, dashed, inner sep=3pt, fit=(Phi)(Plant)(Con1), label=above:{Uncertain System, $G_\phi$}] (frame) {};

\draw[<-]([yshift = +0.2cm, xshift =0.00cm]C.west) -- +(-0.5cm, 0.0cm)node[above, name = r]{$r$};


\draw[->]
    (Plant.east) -- ++(0.4cm,0) node[above] {$x$}          
             |- ++(0,-0.8cm)         
             -| ([xshift=-0.5cm]C.west) 
             -- (C.west);           


\end{tikzpicture}

	
 \caption{Closed loop interconnection with input nonlinearity $\phi$, baseline controller $k_0$ and safety filter $F$.}
 \label{fig:CLIC}
\vspace{-10pt}
\end{figure}

The nonlinearity $\phi:\U \times \R_{\ge 0} \to \R^m$ is assumed to be memoryless but possibly time-varying and can model unknown nonlinear effects \added{such as uncertain actuator gains, saturation, and dead-zones}. We also assume that the nonlinearity is bounded in the sector $[-\theta,\theta]$ with $0<\theta< 1$. This implies that the input $u(t)$ and output $w(t)$ of $\phi$ satisfy the following inequality at each point in time:
\vspaceeq
\begin{align}\label{eq:constraint}
    \| w(t) \|_2 \le \theta \| u(t) \|_2 
    \quad\quad \forall t\ge 0.
\vspaceeq
\end{align}
Thus $\theta\ge 0$ measures the amount of nonlinear perturbation at the plant input. No perturbation corresponds to $\theta=0$, i.e. $w=0$, and is referred to as nominal system $G$ in the remainder of the paper. Systems with more general sector-bounded nonlinearities can be represented in this form via a standard loop-shifting process \cite[Sec. 6.5]{khalil2002nonlinear}. Based on this constraint, we  define the set of feasible outputs of the nonlinearity $\phi$ for a given input $u(t)\in \mcl U$ as:
\vspaceeq
\begin{align}
\mcl{W}(u(t)):=\{ w(t) \, : \, \|w(t)\|_2 \le \theta \|u(t)\|_2\}.
\vspaceeq
\end{align}
The feedback interconnection includes a baseline controller $k_0$ which is designed for reference tracking.
It has access to state-feedback measurements and a reference input vector $r(t) \in \mcl R \subset \R^r$. The latter is confined to the admissible set $\mcl R$. The baseline controller is defined by a Lipschitz continuous function $k_0:\Dx \times \mcl R \to \mcl U$.

We assume this baseline controller provides adequate performance but fails to provide "safe" operation. 
The notion of safety refers to a safe set $\mcl C\subset \Dx$. This set defines the, possibly time-varying, region in the state space in which the system must remain for safe operation \cite{xu2018constrained}. This safe set is defined as the zero-superlevel set of a twice continuously differentiable, time-varying $C^2$ function $h:\Dx \times \R_{\ge 0} \rightarrow \R$:
\vspaceeq
\begin{align}
\label{eq:CBFdef}
\mcl C(t) := \{x \in \Dx \, : \, h(x,t) \geq 0\}. 
\vspaceeq
\end{align}
The boundary and interior of the safe set are denoted by $\partial\mcl C$ and $\text{Int}(\mcl C)$, respectively. We assume $\mcl C$ is non-empty with no isolated points and zero is a regular value of $h$. 

The goal is to design a safety filter $F$ to ensure that the uncertain system $G_\phi$, defined in \eqref{eq:plantG}, operates safely for all nonlinearities in the sector $[-\theta,\theta]$, i.e.,  $x(0) \in \mathcal{C}(0)$ implies $x(t) \in \mathcal{C}(t)$
for all $t \ge 0$. This is equivalent to the forward invariance of $\mcl C(t)$ for the uncertain closed-loop with all possible nonlinearities. Moreover, the safety filter should maintain safety while minimally altering the command from the baseline controller $k_0$. The overall controller, $k$, consists of both the baseline controller $k_0$ and the safety filter $F$ as shown in Fig.~\ref{fig:CLIC}. The closed-loop dynamics with the uncertain system and this combined controller are given by:
\begin{align}
\label{eq:uncertain closed-loop}
\begin{split}
    \dot{x}(t) & = f(x(t)) + g(x(t)) \, \left( u(t) +w(t) \right), \\
    u(t) & = k(x(t),r(t)), \\
    w(t) & = \phi\left( u(t), t \right).
\end{split}
\vspaceeq
\end{align}
We will refer to \eqref{eq:uncertain closed-loop} as nominal closed loop for perturbation levels $\theta=0$, i.e, $w=0$.
The remainder of the section reviews the ingredients required to design a ``safe'' controller.

\subsection{Time-Varying Exponential Control Barrier Functions}\label{ss:tvcbf}
Time-varying exponential control barrier functions (TVCBFs) were introduced in \cite{xu2018constrained}. The nominal closed-loop system is given by
\eqref{eq:uncertain closed-loop} with 
$\theta=0$ and hence $w=0$. TVCBFs ensure safety of this nominal closed loop with respect to time-variant safe sets $\mcl C(t)$ defined by \eqref{eq:CBFdef}. 
We review the formulation for relative degree two. This is a common case due to its relevance for position and rotational dynamics. However, all subsequent derivations generalize to relative degrees other than two, requiring only notational changes.
The following Definition~\ref{defn:TVCI} formalizes time-varying control invariance of a time-varying safe set $\mcl C(t)$  with respect to the nominal closed loop system.
\vspace{0.00in}
\begin{defn}{(\textit{Time-Varying Control Invariance} \cite{xu2018constrained})}
\label{defn:TVCI}
A set $\mcl C(t)$ is time-varying control invariant with respect to the nominal closed loop,
\eqref{eq:uncertain closed-loop} with 
$w=0$,  if there exists a control law $k:\Dx \times \mcl R \to \mcl{U}$ such that the following holds:
\vspaceeq
\begin{align}
x(0)\in \mcl C(0) \Rightarrow 
x(t)\in \mcl C(t) \quad \forall t \ge 0.
\vspaceeq
\end{align}
The nominal closed loop is safe with respect to $\mcl C(t)$ if $\mcl C(t)$ is time-varying control invariant.
\end{defn}
\vspace{0.00in}
A modified Lie derivative of $h(x, t)$ along $f$ is required for time-varying safe sets. It is defined as $\bar{L}_f^i h := (\frac{\partial}{\partial t} + L_f)^i h$ for non-negative integers $i$ \cite{xu2018constrained}. The next definition is a specific case of \cite[Definition~1]{xu2018constrained} for time-varying exponential barrier functions with relative degree two.

\vspace{0.00in}
\begin{defn}{(\textit{Time-Varying Exponential Control Barrier Function})}\label{defn:TVCFB}
Consider the nominal closed-loop system,
\eqref{eq:uncertain closed-loop}, with 
$w=0$. A $C^2$ function $h : \Dx \times \R_{\geq 0} \rightarrow \R$ with a relative degree two is a time-varying control barrier function if there exists an $\alpha > 0$ such that for all $(x, t) \in \Dx \times \R_{\geq 0}$,
\begin{align}
\label{ineq:TVCBF condition with degree 2}
\begin{split}
&\sup_{u \in \mcl U} \, [\bar{L}_f^2 h(x, t) + L_{g} \bar{L}_f h(x, t)u] \\ 
&\geq -2\alpha \bar{L}_f h(x, t) - \alpha^2 h(x, t).
\end{split}
\end{align}
\end{defn}
\vspace{0.02in}
The existence of a TVCBF $h(x)$ implies that there exists a control input $u$ that can render the safe set $\mcl C(t)$ forward invariant and, in particular, prevent trajectories from crossing its boundary $\partial\mcl C(t)$.
 Define the following set of control inputs at a state $x$ and time $t$:
\vspaceeq
\begin{align*}
\mcl U_\text{TVCBF}(x, t) \!:=\! \{{u \!\in\! \mcl U}\! \, : \, \! \, &\bar{L}_f^2 h(x, t) + 2\alpha \bar{L}_f h(x, t) \\ &+ \alpha^2 h(x, t) + L_{g} \bar{L}_f h(x, t)u \geq 0 \}.
\vspaceeq
\end{align*}
Also, define the set $\mcl C_1(t) := \{x \in \Dx : \bar{L}_f h(x, t) + \alpha h(x, t) \geq 0\}$.
The following corollary is an extension of \cite[Proposition~1]{xu2018constrained} for multi-input systems and control systems with reference signals $r$.
\vspace{0.00in}
\begin{cor}
    \label{thm:existence of tvcbf implies forward invariance}
Assume a $C^2$ function $h : \Dx\times \R_{\geq 0} \rightarrow \R$ has relative degree two with respect to the nominal closed loop,
\eqref{eq:uncertain closed-loop} with 
$w=0$. Moreover, assume $h$ satisfies \eqref{ineq:TVCBF condition with degree 2} for some $\alpha > 0$. 
Then, any controller $k: \Dx\times \mcl R \rightarrow \U$ that is Lipschitz continuous in $x$ and $r$, and satisfies $k(x,r)\in\U_\text{TVCBF}(x,t)\,\forall\, (x,r,t) \in \Dx \times \mcl R \times \R_{\ge 0}$ renders the set $\mcl C(t)$ forward invariant provided that $x_0 \in \mcl C(0) \cap \mcl C_1(0)$.
\end{cor}
\begin{proof}
By assumption, $h$ has relative degree two with respect to the nominal system.
Hence $L_{g} h(x, t) = 0$ and $L_{g} \bar{L}_f h(x, t) \neq 0$ for all $(x, t) \in \Dx \times \R_{\geq 0}$. This implies $\bar{L}_f h(x, t) = \dot{h}(x, t)$ and for any $u(x, t) \in \mcl U_\text{TVCBF}(x, t)$ the inequality $\ddot{h}(x, t) + 2\alpha\dot{h}(x, t) + \alpha^2 h(x, t) \geq 0$ is satisfied.  Furthermore, the polynomial $s^2 + 2\alpha s + \alpha^2$ has a repeated root $-\alpha$ with $\alpha > 0$. Since $x_0 \in \mcl C(0) \cap \mcl C_1(0)$, it follows that $h(x_0, 0) \geq 0$ and $\dot{h}(x_0, 0) + \alpha h(x_0, 0) \geq 0$. Thus, Lemma 1 in \cite{xu2018constrained} implies that $x(t) \in \mcl C(t)$ for all $t \geq 0$.
\end{proof}
\vspace{-4pt}
\subsection{Robust Exponential Control Barrier Functions}\label{ss:rcbf}
\vspace{-3pt}
This section provides a short overview of time-invariant robust exponential control barrier functions \cite{buch2021robust}. 
Define a time-invariant $C^2$ function $h : \Dx   \rightarrow \R$ and the set
\vspaceeq
\begin{align}\label{eq:Safeset_TI}
\mcl C := \{x \in \Dx : h(x) \geq 0\}. 
\vspaceeq
\end{align}
The safety of the uncertain closed loop \eqref{eq:uncertain closed-loop} is assured if $x(0) \in \mcl C$ implies that $x(t) \in \mcl C$ for all $t\ge 0$.
The robust control invariance concept proposed in \cite[Definition~4.4]{blanchini2015set} provides a way to ensure safety for uncertain systems. A specific formulation for sector-bounded nonlinearities  $\phi$ was given in
\cite[Definition~4]{buch2021robust}. This definition is repeated next.
\vspace{0.00in}
\begin{defn}{(\textit{Robust Control Invariance}\cite{buch2021robust})}
\label{defn:RCI}
A set $\mcl C \subset \Dx$ is robustly control invariant with respect to the uncertain closed loop \eqref{eq:uncertain closed-loop}, if there exists a control law $k:\Dx \times \mcl R \to \mcl{U}$ such that the following holds for all
admissible nonlinearities $\phi$ in the sector $[-\theta,\theta]$:
\vspaceeq
\begin{align}
x(0)\in \mcl C \Rightarrow 
x(t)\in \mcl C \quad \forall t \ge 0.
\vspaceeq
\end{align}
The uncertain closed loop \eqref{eq:uncertain closed-loop} is robustly safe with respect to $\mcl C$ if $\mcl C$ is robustly control invariant.
\end{defn}
\vspace{0.00in}

Exponential control barrier functions can be used to guarantee safety for the nominal closed loop system, i.e., $\theta = 0$ \cite{ames2019control,nguyen2016exponential}.  Robust exponential control barrier functions guarantee robust control invariance of the uncertain closed-loop system \eqref{eq:uncertain closed-loop}. 
This robust version was proposed in \cite{seiler2021control}. The following definition presents an explicit formulation for an uncertain system of relative degree two. 

\vspace{0.00in}
\begin{defn}{(\textit{Robust Exponential Control Barrier Functions})}
Given a safe set $\mcl C\subset \Dx$ defined by \eqref{eq:Safeset_TI}, a $C^2$ function $h : \Dx  \rightarrow \R$ with a relative degree two is a robust exponential control barrier function for the uncertain system
\eqref{eq:uncertain closed-loop} for all nonlinearities $\phi$ in the sector $[-\theta,\theta]$, if there exists $\alpha > 0$ such that for all $x \in \Dx$,
\begin{align}
\label{ineq:RCBF condition with degree 2}
\begin{split}
&\sup_{u \in \mcl U} \inf_{w \in \mcl W(u)} [L_f^2 h(x) + L_{{g}} L_f h(x)(u + w)] \\ 
&\geq -2\alpha L_f h(x) - \alpha^2 h(x).
\end{split}
\end{align}
\end{defn}
\vspace{0.05in}

\vspace{-6pt}
\section{Robust Time-Varying CBFs}\label{sec:rtvcbf}
\vspace{-4pt}

\subsection{Formulation}\label{ss:form}
In this section, we propose robust time-varying control barrier functions (RTVCBFs), which are designed to provide safety guarantees for the uncertain closed-loop dynamics \eqref{eq:uncertain closed-loop}.
We first formalize the notion of robust time-varying control invariance. This generalizes 
Definitions~\ref{defn:TVCI} and \ref{defn:RCI}.
\vspace{0.00in}
\begin{defn}{(\textit{Robust Time-Varying Control Invariance})}
\label{defn:robust control invariance and robust safety in RTVCBF}
A set $\mcl C(t)$ is robustly time-varying control invariant with respect to the uncertain dynamics \eqref{eq:uncertain closed-loop} if there exists a control law $k:\Dx\times \mcl R \rightarrow \U$ 
such that the following holds for all
admissible nonlinearities $\phi$ in the sector $[-\theta,\theta]$:
\vspaceeq
\begin{align}
x(0)\in \mcl C(0) \Rightarrow 
x(t)\in \mcl C(t) \quad \forall t \ge 0.
\vspaceeq
\end{align}
The system \eqref{eq:uncertain closed-loop} is robustly safe with respect to $\mcl C(t)$ if $\mcl C(t)$ is robust time-varying control invariant.
\end{defn}
\vspace{0.00in}
A robust time-varying exponential control barrier function can be used to formulate safety filters $F$ that guarantee robust safety of uncertain closed loop systems \eqref{eq:uncertain closed-loop} with respect to the time-varying safe sets $\mcl C(t)$ \eqref{eq:CBFdef}. 
The following definition provides an explicit formulation of such control barrier functions for relative degree two. 
\vspace{0.00in}
\begin{defn}{(\textit{Robust Time-Varying Exponential Control Barrier Functions})}
Given a safe set $\mcl C(t) \subset \Dx$ defined by \eqref{eq:CBFdef}, a $C^2$ function $h : \Dx \subset \R^n \times \R_{\geq 0} \rightarrow \R$ with a relative degree two is a robust time-varying control barrier function for the uncertain closed loop \eqref{eq:uncertain closed-loop} for all admissible nonlinearities $\phi$ in the sector $[-\theta,\theta]$, if there exists $\alpha > 0$ such that for all $(x, t) \in \Dx \times \R_{\geq 0}$, 
\begin{align}
\label{ineq:RTVCBF condition with degree 2}
\begin{split}
&\sup_{u \in \mcl U} \inf_{w \in \mcl W(u)} [\bar{L}_f^2 h(x, t) + L_{{g}} \bar{L}_f h(x, t)(u + w)] \\ 
&\geq -2\alpha \bar{L}_f h(x, t) - \alpha^2 h(x, t).
\end{split}
\end{align}
\end{defn}
\vspace{0.00in}
The worst-case behavior of the nonlinearity is represented by a perturbation $w$ that minimizes 
the left-hand side of the inequality \eqref{ineq:RTVCBF condition with degree 2}.
Our goal is to determine a safe control input $u \in \mcl U$ under the corresponding worst-case perturbation $w^*\in \mcl W$. 
The worst-case uncertain input $w^*(u)$, obtained as the solution of the inner optimization problem in \eqref{ineq:RTVCBF condition with degree 2}, is
\begin{align}
\label{eq:worst-case uncertain input with degree 2}
w^*(u) = -\theta \|u\|_2 \cfrac{L_{{g}} \bar{L}_f h(x, t)^{\top}}{\|L_{{g}} \bar{L}_f h(x, t)\|_2}.
\end{align}
Note, that for $\theta \rightarrow 0$ the worst-case disturbance $w^*(u)$ approaches zero and the nominal TVCBF formulation in Definition~\ref{defn:TVCFB} is recovered.
Substituting $w^*(u)$ for $w$ in condition \eqref{ineq:RTVCBF condition with degree 2} provides
\begin{align}
\label{ineq:RTVCBF condition with degree 2 for worst-case uncertain input}
\begin{split}
&\sup_{u \in \mcl U} [\bar{L}_f^2 h(x, t) + L_{\tilde{g}} \bar{L}_f h(x, t)(u + w^*(u)] \\
&\geq -2\alpha \bar{L}_f h(x, t) - \alpha^2 h(x, t).
\end{split}
\end{align}
The constraint \eqref{ineq:RTVCBF condition with degree 2 for worst-case uncertain input} is equivalent to \eqref{ineq:RTVCBF condition with degree 2}. Next, define the set of control inputs 
\begin{align*}
\mcl U_\text{RTVCBF}(x, t) \!:=\! \{{u \in \mcl U} : \, &\bar{L}_f^2 h(x, t) + 2\alpha \bar{L}_f h(x, t) \\ &+ \alpha^2 h(x, t) \\ &+ L_{{g}} \bar{L}_f h(x, t)(u + w^*(u)) \geq 0 \}
\end{align*}
If the supremum in~\eqref{ineq:RTVCBF condition with degree 2 for worst-case uncertain input} is attained, then the above set is non-empty.
The following Theorem 1 
states an existence condition for a robust time-varying control barrier function guaranteeing that the uncertain closed loop \eqref{eq:uncertain closed-loop} is robustly safe with respect to $\mcl C(t)$.

\vspace{0.00in}
\begin{mytheo}
\label{thm:existence of rtvcbf implies robust forward invariance}
Assume a $C^2$ function $h: \Dx \times \R_{\geq 0} \rightarrow \R$ has relative degree two with respect to the uncertain closed loop \eqref{eq:uncertain closed-loop} and satisfies \eqref{ineq:RTVCBF condition with degree 2 for worst-case uncertain input} for some $\alpha > 0$. 
Then, any controller $k: \Dx\times \mcl R \rightarrow \U$ that is Lipschitz continuous in $x$ and $r$, and satisfies $k(x,r)\in\U_\text{RTVCBF}(x,t)\,\forall\, (x,r,t) \in \Dx \times \mcl R \times \R_{\ge 0}$ renders the set $\mcl C(t)$ robust time-varying forward invariant provided that $x_0 \in \mcl C(0) \cap \mcl C_1(0)$.
\end{mytheo}
\begin{proof}
By assumption $h$ has relative degree two with respect to the uncertain closed loop dynamics \eqref{eq:uncertain closed-loop}. Thus $L_{g} h(x, t) = 0$ and $L_{g} \bar{L}_f h(x, t) \neq 0$ for all $(x, t) \in \Dx \subset \R^n \times \R_{\geq 0}$. This implies $\bar{L}_f h(x, t) = \dot{h}(x, t)$ and for any $u(x, t) \in \mcl U_\text{RTVCBF}(x, t)$ the inequality $\ddot{h}(x, t) + 2\alpha\dot{h}(x, t) + \alpha^2 h(x, t) \geq 0$ holds for all inputs $w \in \mcl W$ satisfying \eqref{eq:constraint}. 
Furthermore, the polynomial $s^2 + 2\alpha s + \alpha^2$ has a repeated root $-\alpha$ with $\alpha > 0$. Since $x_0 \in \mcl C(0) \cap \mcl C_1(0)$, it follows that $h(x_0, 0) \geq 0$ and $\dot{h}(x_0, 0) + \alpha h(x_0, 0) \geq 0$. Thus, Lemma 1 in \cite{xu2018constrained} implies that $x(t) \in \mcl C(t)$ for all $t \geq 0$.
\end{proof}
\added{Multiple barriers $\{h_i\}_i$ are handled with one constraint of the form~\eqref{ineq:RTVCBF condition with degree 2 for worst-case uncertain input} for each barrier.  Each constraint shares the same input $u$ but with a specific worst-case signal $w_i^*(u)$. This yields a closed-loop that is safe with respect to each barrier and robust against all possible worst-case signals $w_i^*(u)$.}

\subsection{Control Algorithm}\label{ss:comp}
Based on the approach presented in \cite{buch2021robust}, we propose an efficient online method for computing a safe control input $u$ that minimally interferes with the baseline controller $k_0$. The problem can be posed as the optimization program:
\vspaceeq
\begin{align}
\label{TVCBF-QP initial with degree 2}
\begin{split}
&u^*(x) = \argmin_{u \in \mcl U} \frac{1}{2}\|u - k_0(x,r)\|^2 \\
&\mbox{s.t. } \bar{L}_f^2 h(x, t) + 2\alpha \bar{L}_f h(x, t) + \alpha^2 h(x, t) \\ &\qquad + L_{{g}} \bar{L}_f h(x, t)(u + w^*(u)) \geq 0.
\end{split}
\end{align}
\vspaceeq
Substituting for the the worst-case perturbation $w^*(u)$ as given in Eq. \eqref{eq:worst-case uncertain input with degree 2}) and expanding the cost function provides
\begin{align}
\label{TVCBF-QP final with degree 2}
\begin{split}
&u^*(x) = \argmin_{u \in \mcl U} \bigg[\frac{1}{2}u^{\top}u - {k_0(x,r)}^{\top}u \bigg] \\
&\mbox{s.t. } \bar{L}_f^2 h(x, t) + 2\alpha \bar{L}_f h(x, t) + \alpha^2 h(x, t) \\ &\qquad + L_{{g}} \bar{L}_f h(x, t)u - \theta\|u\|_2 \|L_{{g}} \bar{L}_f h(x, t)\|_2 \geq 0.
\end{split}
\end{align}
For an uncertainty level $\theta=0$, i.e., $w=0$, the nominal system behavior is recovered and \eqref{TVCBF-QP final with degree 2} reduces to the standard time-varying exponential control barrier function quadratic program. For $\theta >0$, the decision variable $u$ enters nonlinearly in the inequality constraint through $\norm{u}_2$. To address this, the optimization can be reformulated into a convex problem in two steps.
First, following \cite{buch2021robust} introduce a slack variable $q$ to rewrite  \eqref{TVCBF-QP final with degree 2} as
\vspaceeq
\vspaceeq
\begin{align}
\label{RTVCBF-SOCP initial with degree 2}
\begin{split}
&\bmtx u^*(x) \\ q^*(x) \emtx= \argmin_{u \in \mcl U, q\in \R_{\ge 0}} [q - {k_0(x,r)}^{\top}u ] \\
&\mbox{s.t. } \bar{L}_f^2 h(x, t) + 2\alpha \bar{L}_f h(x, t) + \alpha^2 h(x, t) \\ &\qquad + L_{{g}} \bar{L}_f h(x, t)u \geq \theta \|L_{{g}} \bar{L}_f h(x, t)\|_2\|u\|_2 \\
&\qquad  2q \geq \|u\|_2^2.
\end{split}
\vspaceeq
\end{align}
The reformulated problem is a minimization over $u \in \mcl U$ and $q \in \R_{\ge 0}$ with $2q^* = u^{*^{\top}}u^*$. Finally, the second constraint is recast as a rotated second-order cone (SOC) condition following \cite[Sec. 10.1]{calafiore2014optimization}.
This gives the Robust TVCBF-second-order-cone-program (Robust TVCBF-SOCP):
\begin{align}
\label{RTVCBF-SOCP final with degree 2}
\begin{split}
&\bmtx u^*(x) \\ q^*(x) \emtx= \argmin_{u \in \mcl U, q\in \R_{\ge 0}} [q - {k_0(x,r)}^{\top}u ] \\
&\mbox{s.t. } \theta \|L_{{g}} \bar{L}_f h(x, t)\|_2\|u\|_2 \leq \bar{L}_f^2 h(x, t) \\
&\qquad + 2\alpha \bar{L}_f h(x, t) + \alpha^2 h(x, t) + L_{{g}} \bar{L}_f h(x, t)u\\
&\qquad  \left\|\bsmtx \sqrt{2}u \\ q - 1 \esmtx \right\|_2 \leq q + 1.
\end{split}
\end{align}
This can be efficiently solved online. \added{Note that the approach naturally extends to multiple barrier functions by adding the respective linear inequality conditions.}

\subsection{Feasibility}\label{ss:feas}
The feasibility of the optimization \eqref{RTVCBF-SOCP final with degree 2} is essential to guarantee robust safety of the control system. We consider two cases: 1) unconstrained control inputs and 2) Euclidean norm bounded
control inputs. The latter appear in problems assuming maximum torque and force magnitude bounds, e.g., in thrust-vectoring control of space launchers or spacecraft.

First, we consider the case of unconstrained control inputs. To simplify the notation,  we define
\begin{equation*}
\begin{split}
    c_1(x, t) &:= \bar{L}_f^2 h(x, t) + 2\alpha \bar{L}_f h(x, t) + \alpha^2 h(x, t)\\
    c_2(x,t)&:=L_{{g}} \bar{L}_f h(x, t). 
\end{split}
\end{equation*}
The main constraint in \eqref{RTVCBF-SOCP final with degree 2} can be rewritten in the following simpler form based on this notation:
\begin{align*}
\label{RTVCBF-SOCP-Simple}
    \theta \|c_2(x,t) \|_2\|u\|_2 \leq 
    c_1(x,t) + c_2(x,t) u
\end{align*}
The next theorem formalizes the feasibility analysis of the RTVCBF-SOCP for the case of unconstrained control inputs.

\vspace{0.00in}
\begin{mytheo}\label{thm:FeasUnconst}
Assume a $C^2$ function $h: \Dx \times \R_{\geq 0} \rightarrow \R$ has relative degree two with respect to the uncertain system dynamics \eqref{eq:uncertain closed-loop}. If $\theta \in [0,1)$ and the control input is unconstrained then  the RTVCBF-SOCP problem is feasible for all $(x, t) \in \Dx \times \R_{\geq 0}$.
\end{mytheo}
\begin{proof}
The optimization \eqref{RTVCBF-SOCP initial with degree 2} and the RTVCBF-SOCP \eqref{RTVCBF-SOCP final with degree 2} are equivalent. Hence we consider \eqref{RTVCBF-SOCP initial with degree 2} in the remainder of the proof. Since $h$ has relative degree two with respect to the uncertain dynamics \eqref{eq:uncertain closed-loop}, $c_2(x,t) \neq 0\,\forall\, (x, t) \in \Dx \times \R_{\geq 0}$. 

If $c_1(x, t) \geq 0$ then select $u(t) := 0 \in \R^m$ and $q(t) := 0$. The first constraint in \eqref{RTVCBF-SOCP initial with degree 2} simplifies to  $c_1(x,t)\ge 0$ which is satisfied by assumption.  The second constraint 
in \eqref{RTVCBF-SOCP initial with degree 2}
holds trivially. Thus, $[ u(t)^\top,q(t)]^\top$ is a feasible solution for \eqref{RTVCBF-SOCP initial with degree 2}. 

If $c_1(x, t) < 0$ then select
\begin{equation}
\label{eq:uqFeas}
\begin{split}
    &u(t) := \bigg(-\frac{c_1(x, t)}{(1 - \theta)\|c_2(x,t)\|_2^2}\bigg)c_2(x,t)^{\top} \in \R^m \\
&q(t) := \frac{1}{2} \|u\|_2^2
\end{split}
\end{equation}
This choice of $u(t)$ (combined with $0<\theta<1$) yields
\begin{align*}
&c_1(x, t) + c_2(x,t)u = -\frac{\theta}{1-\theta} c_1(x,t) \\
& \theta \|c_2(x,t)\|_2 \|u \|_2
= \frac{\theta}{1-\theta} |c_1(x,t)|
\end{align*}
Since $c_1(x,t)<0$ for this case, it follows that
\begin{align*}
&c_1(x, t) + c_2(x,t)u = \theta \|c_2(x,t)\|_2\|u\|_2
\end{align*}
Thus, $[ u(t)^\top,q(t)]^\top$ is a feasible solution for \eqref{RTVCBF-SOCP initial with degree 2}. 
\end{proof}

\vspace{0.00in}

Next, we  consider the case of
control inputs with a Euclidean norm bounded by $u_{\max} > 0$, i.e., $\norm{u(t)}_2\le u_\text{max}$. The feasibility of the RTVCBF-SOCP under this constraint is established in the next theorem.

\vspace{0.00in}
\begin{mytheo}\label{thm:FeasEucl}
Assume a $C^2$ function $h: \Dx \times \R_{\ge 0} \rightarrow \R$ has relative degree two with respect to the uncertain dynamics \eqref{eq:uncertain closed-loop}. If $\theta \in [0,1)$ and the Euclidean norm of the control input $u(t)$ is bounded by $u_\text{max} > 0$ then the RTVCBF-SOCP problem is feasible for $(x, t) \in \Dx \times \R_{\geq 0}$ if and only if 
\begin{align}
\label{eq:umaxFeasCond}
u_{\max} \geq -\frac{c_1(x, t)}{(1 - \theta)\|c_2(x,t)\|_2}
\end{align}
\end{mytheo}
\vspace{0.05in}
\begin{proof}
The proof is similar to that given for Theorem~\ref{thm:FeasUnconst}.  Note that $u(t)=0$ is feasible when $c_1(x,t)\ge 0$.  Hence it suffices to only consider the case when $c_1(x, t) < 0$.

\noindent$(\Leftarrow)$
Assume Equation~\eqref{eq:umaxFeasCond} holds.
Then select $u(t)$ and $q(t)$ as in 
\eqref{eq:uqFeas}. It was shown in the proof
for Theorem~\ref{thm:FeasUnconst} that this choice
is feasible. Moreover, it follows
directly from \eqref{eq:umaxFeasCond} that this choice for the control input satisfies
$\| u(t)\|_2 \le u_{max}$.

\noindent$(\Rightarrow)$
Assume Equation~\ref{eq:umaxFeasCond} is violated.
Then for any $u(t)$ with $\norm{u(t)}_2\le u_\text{max}$, we have
\begin{align*}
& c_2(x,t)u(t) -\theta \|c_2(x,t)\|_2 \|u(t)\|_2
\\
& \quad 
\le (1-\theta) \|c_2(x,t)\|_2 \|u(t)\|_2 
< -c_1(x,t)
\end{align*}
The last inequality uses
\begin{align*}
\|u(t)\|_2 \leq u_{\max} < -\cfrac{c_1(x, t)}{(1 - \theta)\|c_2(x,t)\|_2}.
\end{align*}
Hence, for any $\|u(t)\|\le u_{max}$,
\begin{align*}
    c_1(x,t) + c_2(x,t) u <
    \theta \|c_2(x,t) \|_2\|u\|_2. 
\end{align*}
Thus, the RTVCBF-SOCP constraint is violated.
\end{proof}

The feasibility condition in Theorem~\ref{thm:FeasEucl} is necessary
and sufficient for a single barrier constraint and multiple inputs. The following remark addresses the case of multiple barrier functions.
\begin{remark}\label{rem:multibarrier}
In a multi-barrier scenario, all constraints share the same
control input $u$. Thus, satisfaction of the scalar feasibility condition for each barrier
individually is necessary but not sufficient for joint feasibility. In other words, the joint feasibility problem can be rendered infeasible if different barriers require incompatible control directions.
A sufficient condition for joint feasibility can be obtained under an alignment assumption on the barrier constraints. Let
\begin{equation*}
    d_i(x,t) := \frac{c_{2,i}(x,t)^\top}{\norm{c_{2,i}(x,t)}_2}
\end{equation*}
denote the unit control direction associated with barrier $h_i$, where
\begin{equation*}
c_{2,i}(x,t) := L_g \bar{L}_f h_i(x,t). 
\end{equation*}
If there exists a common unit 
vector
$v(x,t)$ such that
\begin{align}\label{eq:coneAlign}
  v(x,t)^\top d_i(x,t) > \theta
\end{align}
for all barrier constraints $i$, then the single control direction $u = \rho\,v(x,t)$ satisfies every barrier
constraint for sufficiently large $\rho > 0$. For every constraint requiring control action, i.e., $c_{1,i}(x,t)<0$, the required magnitude satisfies
\begin{equation*}
    \rho \ge \frac{-c_{1,i}(x,t)}{\norm{c_{2,i}(x,t)}_2(v(x,t)^\top d_i(x,t)-\theta)}.
\end{equation*}
If~$c_{1,i}(x,t) \ge 0$ for all $i$, then $u=0$ is jointly feasible. Otherwise, joint feasibility is guaranteed if
\begin{equation}\label{eq:umaxjoint}
    u_\text{max} \ge\!\! \max_{i:c_{1,i}(x,t)<0}\!\! \frac{-c_{1,i}(x,t)}{\norm{c_{2,i}(x,t)}_2(v(x,t)^\top d_i(x,t) -\theta)}.
\end{equation}
The collinear case $d_i(x,t)=v(x,t)$ for all $i$ is the special instance
in which the condition~\eqref{eq:umaxjoint} reduces to
\begin{align}\label{eq:multiFeasCond}
  u_\text{max} \geq \max_{i:c_{1,i}(x,t)<0}
  \frac{-c_{1,i}(x,t)}{(1-\theta)\norm{c_{2,i}(x,t)}_2},
\end{align}
i.e., the maximum of the individual bounds in Theorem~\ref{thm:FeasEucl}. In this case, $v(x,t)$ additionally maximizes the robust margin of every barrier.
\end{remark}

\vspace{-2pt}
\section{Numerical Example}\label{sec:exmp}
\vspace{-6pt}

The example considers a planar docking maneuver of a CubeSat with a decommissioned spacecraft in low earth orbit (LEO). This is an essential maneuver in debris removal missions using CubeSats and requires safe and robust autonomy.

\vspace{-6pt}
\subsection{Satellite Dynamics}

The CubeSat's docking dynamics are formulated with respect to a local vertical local horizontal (LVLH) coordinate frame fixed to the target and defined following \cite{Alfriend2010}. These dynamics have the form of~\eqref{eq:plantG} with
\begin{equation}\label{eq:SatDyn}
\begin{split}    
    f(x(t)) &= \bsmtx v_x\\ v_y \\\!\! 2 \omega(t) v_y + \dot{\omega}(t) p_y + \omega(t)^2p_x + \frac{\mu}{r_\text{CS}(t)^2} - \mu \frac{r_\text{T}(t) + p_x}{r_\text{CS}(t,x)^3} \!\!\\
    -2\omega(t) v_x - \dot{\omega}(t) p_x + \omega(t)^2p_y - \mu \frac{p_y}{r_\text{CS}(t,x)^3} \esmtx \\g(x(t)) &= \bsmtx 0 & 0 & \frac{1}{m_\text{CS}}&0\\ 0 & 0 & 0 & \frac{1}{m_\text{CS}} \!\!\esmtx^\top.
\end{split}
\end{equation}
The state vector $x = [p_x,p_y,v_x,v_y]^\top$ is comprised of 
the relative positions ($p$) and velocities ($v$) of the CubeSat radial to ($x$) and in direction ($y$) of the target velocity, respectively. The control input
$u = [T_x,T_y]^\top$ is the thrust in $x$ and $y$ direction. The CubeSat is modeled after the OAAN CubeSat \cite{Pei2017} with mass $m_\text{CS} = 4$ kg. Four cold gas thrusters provide $\pm 300$ mN of thrust $T_x$ and $T_y$. The maximum simultaneous thrust $T$ is limited, due to power constraints, by $T_x^2 + T_y^2 \le (300\mbox{mN})^2$. The nonlinearity $\phi$ is bounded to the sector $\pm\theta= 0.75$, equivalent to $75\%$ maximum plant input uncertainty, to account for large thrust uncertainties for small satellites.

In~\eqref{eq:SatDyn}, $r_\text{CS} = \sqrt{(r_\text{T} + p_x)^2 + p_y^2}$ is the CubeSat's and $r_\text{T}$ the target's distance to the center of the central body, respectively. The LVLH rotation rate and angular acceleration are denoted by $\omega$ and $\dot{\omega}$, respectively. The orbit-coefficient time histories are calculated for ICESat-2 (NORAD 43613) using the publicly available two-line element sets \cite{icesat2_tle}, propagated using a two-body Kepler propagation. 
The gravitational parameter $\mu$ has a value of $3.986 \cdot10^{14}\,\text{m}^3/\text{s}^2$. 
\vspace{-6pt}
\subsection{Docking Scenario}
\vspace{-2pt}
The maneuver starts at $p_x = -10$ m and $p_y = 0$ m and $v_x = v_y =0$, i.e., with zero initial velocity. The initial target docking position is located at $p_{\text{f},x} = 0$ m and $p_{\text{f},y} = 0$. 
During the approach, the CubeSat identifies a cluster of space debris with a radius of $r_{0} = 0.35$ m at $p = [-5\,\text{m},0\,\text{m}]= [c_{x_0},c_{y_0}]$ that drifts with a constant velocity $v_{\text{c},x} = 1$ mm/s and $v_{\text{c},y} = 2$ mm/s. More objects are identified while the CubeSat gets closer to the target which increases the cluster size by $s_{r} =1$ mm/s. Note that $r_0$ includes a $0.05$ m safety margin to account for the CubeSat's cross-section. While avoiding the obstacle, the CubeSat must also satisfy a line-of sight (LoS) constraint to maintain visual contact with the docking location. This constrain is modeled as a cone centered at $p_\text{f}$, with half-cone angle $\gamma = 5$ deg and width of $0.1$ m at $p_{\text{f}}$. The cone moves with a constant $y$ velocity $v_{\text{LoS},y} =-1$ mm/s as the CubeSat updates the docking position $p_{\text{f}}$. As the derivations in Section~\ref{sec:rtvcbf} generalize directly to multiple barrier functions $h_i$,  $i\in \{1,2,\ldots, n\}$, we can, according to the two safety requirements, define the safe set as in~\eqref{eq:CBFdef} via
\vspaceeq
\begin{align}
    h_1 &= (p_x-\bar{c}_x(t))^2 + (p_y-\bar{c}_y(t))^2 -r_{\text{c}}(t)^2\\
    h_2 &= \tan{(\gamma)} \,p_x + n_1(t) - p_y\\
    h_3 &= p_y -\tan{(-\gamma)} \,p_x -n_2(t),
\vspaceeq
\end{align}
where $\bar{c}_x(t) = c_{x_0} + v_{\text{c},x} t$, $\bar{c}_y(t) = c_{y_0} + v_{\text{c},y} t$, $r_\text{c} = r_{0} + s_rt$ $n_i = n_{i,0} + v_{\text{LoS},y}$ with  $i=2,3$. 
The relevant terms to compute $\bar{L}_f h_i(x, t)$, $\bar{L}_f^2 h_i(x, t)$ and $L_g\bar{L}_f h_i(x, t)$ follow the descriptions in Section~\ref{sec:rtvcbf}.  
\replaced{Both, the TVCBF and RTVCBF use $\alpha = 0.5$.}{The TVCBF and RTVCBF place the repeated design poles at $s=-\alpha =0.5$.} The TVCBF control signal is calculated using Matlab's \texttt{quadprog}. The algorithm \texttt{coneprog} is used in case of the RTVCB-SOCP.

The CubeSat shall approach the target with a constant velocity of $0.03$ m/s and follow the $y$ component of the docking location $p_\text{f}$. Thus, a linear quadratic regulator (LQR), $K$, is designed for the linearized CubeSat dynamics. The cost matrices are chosen as $Q = \mbox{diag}(10,10^3, 10^8, 10^5)$ and $R = \mbox{diag}(10^7,10^7)$ yielding $K = \bsmtx 0.001 & 0 & 3.164 & 0\\ 0 & 0.010 & 0 & 0.300\esmtx$. The LQR controller follows the reference command 
$r = \bsmtx 0 & \frac{n_1(t) +n_2(t)}{2} & 0.03 & 0 \esmtx$ using the control law $u_0(r,x) = K\cdot(r-x)$. 
We implement three control architectures following Fig. \ref{fig:CLIC}: 1) baseline LQR controller, 2) baseline LQR with nominal TVECBF safety filter, and 3) baseline LQR with the novel RTVECBF safety filter. Architecture 3) also checks the feasibility condition in Theorem 3. For guaranteed infeasibility, it skips the SOCP and commands the maximum control effort.

\vspace{-3pt}
\subsection{Results}
\vspace{-3pt}
The docking trajectories using the worst-case perturbation $w^*$ corresponding to $h_1$ \added{as derived in Section~\ref{ss:form}} are shown in the top plot of Fig.~\ref{fig:LatMotion}. 
Pure LQR control (\ref{pl:LQR}) cannot evade the obstacle. The safety filter using the nominal TVCBF (\ref{pl:TVBF}) performs an evasive maneuver, but violates the safety criterion at $t = 152$ s. The CubeSat crosses the debris field at this time instance (\ref{pl:debris2}) which moved from its initial location (\ref{pl:debris1}) at $t=0$ s. This is visualized by the negative distance to the safety boundary shown in the bottom plot of Fig.~\ref{fig:LatMotion} \added{, which is directly related to $h_1$}.
The novel safety filter based on the RTVCBF (\ref{pl:RBTV}) safely avoids the obstacle. This is achieved by choosing a more cautious and slower trajectory by applying more thrust that account for the system uncertainty as shown in the middle plot of Fig.~\ref{fig:LatMotion}. \added{The minimum, maximum, and mean solver time on a M4 MacBook are $0.0010$ s, $0.0024$ s, and $0.0014$ s, respectively.}

\begin{figure}[ht!]
\centering
\begin{tikzpicture}
\definecolor{blue1}{RGB}{222,235,247}
\definecolor{blue2}{RGB}{158,202,225}
\definecolor{blue3}{RGB}{49,130,189}
%

\begin{groupplot}[group style={
                      	group name=myplot,
                      	group size= 1 by 4,
                        vertical sep=0.55cm, 
                        horizontal sep = 1.75cm},
                      	height=0.4\columnwidth,
                      	width = 0.9\columnwidth,
                      	xmajorgrids=true,
			ymajorgrids=true,
			 grid style={densely dotted,white!60!black},
			  xmin = -20, xmax = 20,
			  ymin = -5, ymax = 5,
              legend style={at={(+0.66,0.43)},anchor=west},
			   ]]

\nextgroupplot[	  ylabel= {$p_y\, [\text{m}]$},
				  xlabel= {},
                  axis equal,
                  xmin = -6, xmax = 0,
			  ymin = -0.5, ymax = 0.5,
    declare function={
        m1 = -0.1763;
        n1 = -0.2722;
        f1(\x) = m1*\x + n1;           
        m2 = 0.1763;
        n2 = -0.2722-0.1;
        f2(\x) = m2*\x + n2;
        }
                 ]
\addplot[TealBlue, line width = 0.5] table[x expr = \thisrowno{1} ,y expr = \thisrowno{2} ,col sep=comma] {figures/TrajSat.csv};\label{pl:LQR}
\addplot[YellowOrange, line width = 0.5] table[x expr = \thisrowno{3} ,y expr = \thisrowno{4} ,col sep=comma] {figures/TrajSat.csv};\label{pl:TVBF}
\addplot[Fuchsia, line width = 0.5] table[x expr = \thisrowno{5} ,y expr = \thisrowno{6} ,col sep=comma] {figures/TrajSat.csv};\label{pl:RBTV}


\addplot+[draw=gray, mark = x, mark size=1.5pt, mark options={fill=gray}] coordinates {(-5,-0.1)} circle[radius=0.3] 
    node[pos=0.0, above right, yshift = 0.15cm, xshift = 0.2cm] {\textcolor{gray}{$t=0\,\text{s}$}};\label{pl:debris1}

\addplot+[draw=gray, mark = square*, mark size=1.5pt, mark options={fill=gray}, dashed] coordinates {(-4.8477, -0.4046)} circle[radius=0.4523] 
    node[pos=0.0, above right, yshift = -0.4cm, xshift = 0.5cm] {\textcolor{gray}{$t=152\, \text{s}$}};\label{pl:debris2}

\draw[<-, gray] (axis cs:-4.8477, -0.4046) -- (axis cs:-5,-0.1)
    node[midway, right] {};


\addplot [domain=-10:0, samples=2, color=gray] {f1(x)};
\addplot [domain=-10:0, samples=2, color=gray] {f2(x)};


\nextgroupplot[	  ylabel= {$v_x\, [\text{cm/s}]$},
				  xlabel= {},
                  xmin = -6, xmax = 0,
			  ymin = -0.0, ymax = 4,
                 ]

\addplot[TealBlue, line width = 0.5] table[x expr = \thisrowno{1} ,y expr = \thisrowno{7}*100 ,col sep=comma] {figures/TrajSat.csv};\label{pl:LQR}
\addplot[YellowOrange, line width = 0.5] table[x expr = \thisrowno{3} ,y expr = \thisrowno{9}*100 ,col sep=comma] {figures/TrajSat.csv};\label{pl:TVBF}
\addplot[Fuchsia, line width = 0.5] table[x expr = \thisrowno{5} ,y expr = \thisrowno{11}*100 ,col sep=comma] {figures/TrajSat.csv};\label{pl:RBTV}


%


\nextgroupplot[	  ylabel= {$\Delta r \, [\text{m}]$},
				  xlabel= {$p_x\, [\text{m}]$},
			  	ymin = -0.5, ymax = 0.5,
				 xmin = -6, xmax = -4,
                 ]

\addplot[YellowOrange, line width = 1.0] table[x expr = \thisrowno{2} ,y expr = \thisrowno{5} ,col sep=comma] {figures/hbfSat.csv};\label{pl:TVBF}
\addplot[Fuchsia, line width = 1.0] table[x expr = \thisrowno{3} ,y expr = \thisrowno{6} ,col sep=comma] {figures/hbfSat.csv};\label{pl:RBTV}
\end{groupplot}

\begin{axis}[
    at={(myplot c1r2.south west)},
    anchor=south west,
    width=0.9\columnwidth,
    height=0.4\columnwidth,
    xmin=-6, xmax=0,
    ymin=0, ymax=200,
    axis x line=none,
    axis y line*=right,
    ytick pos=right,
    yticklabel pos=left,
    ytick align=inside,
    ytick={100},
    yticklabels={100},
    ylabel={$\norm{T}\, [\text{mN}]$ (- -)},
    ylabel style={text=gray!70!black, xshift=0pt},
    yticklabel style={text=gray!70!black, anchor=east, xshift=-3pt},
    ymajorgrids=false,
    xmajorgrids=false,
    grid=none,
    background/.style={fill=none}
]




\addplot[TealBlue, dashed, line width=0.8pt, opacity=0.8]
table[x expr=\thisrowno{1}, y expr=\thisrowno{2}*1000, col sep=comma] {figures/uctrlSat.csv};

\addplot[YellowOrange, dashed, line width=0.8pt, opacity=0.8]
table[x expr=\thisrowno{3}, y expr=\thisrowno{4}*1000, col sep=comma] {figures/uctrlSat.csv};

\addplot[Fuchsia, dashed, line width=0.8pt, opacity=0.8]
table[x expr=\thisrowno{5}, y expr=\thisrowno{6}*1000, col sep=comma] {figures/uctrlSat.csv};

\end{axis}

\end{tikzpicture}
\vspace{-5pt}
\caption{Simulation results: LQR (\ref{pl:LQR}), TVCBF (\ref{pl:TVBF}), RTVCBF (\ref{pl:RBTV})}
    \label{fig:LatMotion}
\vspace{-10pt}
\end{figure}

\begin{figure}[ht!]
\centering
\begin{tikzpicture}
\definecolor{blue1}{RGB}{222,235,247}
\definecolor{blue2}{RGB}{158,202,225}
\definecolor{blue3}{RGB}{49,130,189}
%

\begin{groupplot}[group style={
                      	group name=myplot,
                      	group size= 1 by 2,
                        vertical sep=0.55cm, 
                        horizontal sep = 1.75cm},
                      	height=0.4\columnwidth,
                      	width = 0.9\columnwidth,
                      	xmajorgrids=true,
			ymajorgrids=true,
			 grid style={densely dotted,white!60!black},
			  xmin = -20, xmax = 20,
			  ymin = -5, ymax = 5,
              legend style={at={(+0.66,0.43)},anchor=west},
			   ]]

\nextgroupplot[	  ylabel={$\norm{T}\, [\text{mN}]$},
				  xlabel= {},
                  xmin = -6, xmax = -4,
			  ymin = 0, ymax = 200,
                 ]

\addplot[ForestGreen, dashed, line width=0.8pt]
table[x expr=\thisrowno{3}, y expr=\thisrowno{4}*1000, col sep=comma] {figures/uctrlSat030.csv};
\addplot[ForestGreen, line width=0.8pt]
table[x expr=\thisrowno{5}, y expr=\thisrowno{6}*1000, col sep=comma] {figures/uctrlSat030.csv};\label{pl:030}

\addplot[BlueViolet, dashed, line width=0.8pt]
table[x expr=\thisrowno{3}, y expr=\thisrowno{4}*1000, col sep=comma] {figures/uctrlSat050.csv};
\addplot[BlueViolet, line width=0.8pt]
table[x expr=\thisrowno{5}, y expr=\thisrowno{6}*1000, col sep=comma] {figures/uctrlSat050.csv};\label{pl:050}

\addplot[BrickRed, dashed, line width=0.8pt]
table[x expr=\thisrowno{3}, y expr=\thisrowno{4}*1000, col sep=comma] {figures/uctrlSat070finer.csv};
\addplot[BrickRed, line width=0.8pt]
table[x expr=\thisrowno{5}, y expr=\thisrowno{6}*1000, col sep=comma] {figures/uctrlSat070finer.csv};\label{pl:070}

\addplot[Goldenrod, dashed, line width=0.8pt]
table[x expr=\thisrowno{3}, y expr=\thisrowno{4}*1000, col sep=comma] {figures/uctrlSat090finer.csv};
\addplot[Goldenrod, line width=0.8pt]
table[x expr=\thisrowno{5}, y expr=\thisrowno{6}*1000, col sep=comma] {figures/uctrlSat090finer.csv};\label{pl:090}


%


\nextgroupplot[	  ylabel= {$\Delta r \, [\text{m}]$},
				  xlabel= {$p_x\, [\text{m}]$},
			  	ymin = -0.5, ymax = 0.5,
				 xmin = -6, xmax = -4,
                 ]


\addplot[ForestGreen, dashed, line width = 1.0] table[x expr = \thisrowno{2} ,y expr = \thisrowno{5} ,col sep=comma] {figures/hbfSat030.csv};
\addplot[ForestGreen, line width = 1.0] table[x expr = \thisrowno{3} ,y expr = \thisrowno{6} ,col sep=comma] {figures/hbfSat030.csv};

\addplot[BlueViolet, dashed, line width = 1.0] table[x expr = \thisrowno{2} ,y expr = \thisrowno{5} ,col sep=comma] {figures/hbfSat050.csv};
\addplot[BlueViolet, line width = 1.0] table[x expr = \thisrowno{3} ,y expr = \thisrowno{6} ,col sep=comma] {figures/hbfSat050.csv};\label{pl:RBTV}

\addplot[BrickRed, dashed, line width = 1.0] table[x expr = \thisrowno{2} ,y expr = \thisrowno{5} ,col sep=comma] {figures/hbfSat070finer.csv};
\addplot[BrickRed, line width = 1.0] table[x expr = \thisrowno{3} ,y expr = \thisrowno{6} ,col sep=comma] {figures/hbfSat070finer.csv};

\addplot[Goldenrod, dashed, line width = 1.0] table[x expr = \thisrowno{2} ,y expr = \thisrowno{5} ,col sep=comma] {figures/hbfSat090finer.csv};
\addplot[Goldenrod, line width = 1.0] table[x expr = \thisrowno{3} ,y expr = \thisrowno{6} ,col sep=comma] {figures/hbfSat090finer.csv};

\end{groupplot}

%
%
%
%
%
%
%
%
%
%

\end{tikzpicture}
\vspace{-5pt}
\caption{Comparison of RTVCBF (solid) and TVCBF (dashed) for varying uncertainty levels: $\theta = 0.3$ (\ref{pl:030}), $\theta = 0.5$ (\ref{pl:050}), $\theta = 0.7$ (\ref{pl:070}), and $\theta = 0.9$ (\ref{pl:090})}
    \label{fig:StudyTheta}
\vspace{-10pt}
\end{figure}
We now investigate the effect of different uncertainty levels on the CubeSat trajectory. We focus on the distance to the safety boundary $\Delta r$ defined by $h_1$ and the total thrust magnitude $\norm{T}$ required for the evasive maneuver. The former was identified as the most critical constraint in extensive tests and also presents the most safety-critical constraint in the docking scenario. The latter directly relates to the feasibility condition and provides insights into how safety is guaranteed.
The nominal CBF approach fails to provide safety for $\theta$ larger than $0.25$. In other words, the CubeSat traverses the debris field. The robust approach provides safety up to an uncertainty norm bound of $0.9$. For larger values, the SOCP feasibility check fails in the proximity of the debris, i.e., there exists no valid control signal for the given ball-constraint, and the CubeSat traverses the unsafe region. 

Fig.\ref{fig:StudyTheta} shows numerical values of $\Delta r$ and $\norm{T}$ for uncertainty bounds $\theta \in \{0.3, 0.5, 0.7, 0.9\}$. It can be seen that the robust approach avoids the unsafe region through earlier and larger control commands. The rapid growth of $\norm{T}$ seen in the top plot of Fig.~\ref{fig:StudyTheta} is consistent with the $\frac{1}{1-\theta}$ scaling of the required $u_\text{max}$ in Theorem~\ref{thm:FeasEucl} for $\theta \rightarrow 1^-$.
The line-of-sight constraint was not close to being violated in any of the investigated scenarios. Overall, the robust RTVCBF approach significantly improves  safety in the presence of input uncertainty compared to the standard TVCBF approach.

\section{CONCLUSION}
\vspace{-6pt}
The paper introduced an approach for robust time-varying control barrier functions that accounts for time-varying safe sets and plant input nonlinearities. The safety filter is formulated as the solution of a second-order cone program. A \added {novel online} feasibility \added{test} is derived for this optimization. The approach was demonstrated on a CubeSat docking scenario. Future work will further investigate feasibility conditions under different assumptions on the input constraints.

\bibliographystyle{IEEEtran}
\bibliography{ReferencesPaper}

\end{document}